\documentclass{acta}

\usepackage{epsfig}
\usepackage{supertabular,lscape,epsfig}

\usepackage{graphicx}	
\usepackage{amsmath}	
\usepackage{amssymb}	
\usepackage{gensymb}
\usepackage{multirow}
\usepackage{multicol}
\usepackage{siunitx} 

\SetPages{0}{0}

\SetVol{00}{2018}

\begin{document}

\begin{Titlepage}
\Title{Secular period variations of RR Lyrae stars in NGC 4147}

\Author{A.~~L~u~n~a\\ and \\ A.~~A~r~e~l~l~a~n~o~~F~e~r~r~o}{Instituto de Astronom\'ia, Universidad Nacional Aut\'onoma de M\'exico, Ciudad Universitaria, 04510, M\'exico.\\
e-mail:aluna@astro.unam.mx}

\Received{Month Day, Year}
\end{Titlepage}

\Abstract{
A compilation of photometric time-series data of NGC 4147 spanning 58 years, enabled to study the secular behaviour of the pulsation period
in the population of RR Lyrae stars in the cluster. The traditional $O-C$ diagram approach was employed. Three stars display a significant period decrease and two a comparable period increase. Nevertheless, the median value of the whole sample is nearly zero, which seems consistent with theoretical expectations. The limitations of empirical determinations of secular period changes in globular clusters to corroborate evolutionary theoretical predictions are discussed.}{globular clusters: individual: NGC 4147 -- stars: variables: RR Lyrae}

\section{Introduction}
The observed secular period changes in RR Lyrae stars (RRLs) in Globular Clusters have been discussed in literature for the last few decades. However, it is not clear if they are caused by the evolutionary changes or are merely events of stochastic nature (e.g. Bal\'azs-Detre \& Detre, 1965; Sweigart \& Renzini, 1979). The comparison of evolutionary models (Lee, 1991; Catelan et al., 2004) with empirically calculated period change rates, is marked by a number of loose threads that we aim to highlight and discuss in the present work. 

It is known that at horizontal branch (HB) evolutionary stage, a star burns its helium core into carbon and oxygen, the temperature and luminosity gradually change as do the mean radius and density. If the density of a pulsating star increases, a decrease in the period should be expected and viceversa. Given the helium burning lifetime, evolutionary period change rates are however expected to be generally smaller than 0.1 day per million years and to remain nearly constant over a century or so (Dorman, 1992). The observational challenge is to determine such small period change rates. The traditional method to study the period variations is the $O-C$ diagram (e.g. Arellano Ferro et al., 2016, 2018b). The success of the method rests on two conditions: the time series should not have long gaps and be as long as possible, not to mention that the quality of the light curves should be reasonably good to estimate the time of maximum brightness with the desired accuracy.

Studies of period variations of RRLs in globular clusters with data covering more than 60 years are scarce; for example, M3 (Corwin \& Carney, 2001; Jurcsik et al., 2012), M5 (Arellano Ferro et al., 2016; Szeidl et al., 2011), NGC 6934 (Stagg \& Wehlau, 1980), M14 (Wehlau \& Froelich, 1994), M15 (Silberman \& Smith, 1995), NGC 7006 (Wehlau et al., 1999), $\omega$ Cen (Jurcsik et al., 2001), and M107 (NGC 6171) (Arellano Ferro et al., 2018b). In the present investigation, we concentrate on NGC 4147, for which we have been able to compile data for about 58 years. We shall discuss our results and present our conclusions, their limitations and caveats in the theoretical and observational framework.

The paper is organized as follows: \S 2 describes the available data for the RRLs in NGC 4147 since 1954. \S 3 outlines the $O-C$ method. In \S 4 the $O-C$ diagrams of all known RRLs in the cluster are presented and the corresponding period change rates $\beta$ are calculated. The highlights on each individual star are given. In \S 5 the mean period variations found in NGC 4147 are put into the perspective of the theoretical predictions and the empirical results in other clusters.
Finally in \S 6 we summarize our conclusions.

\section{The NGC 4147 data}

The globular cluster NGC 4147 is located at 19.3 kpc from the Sun and has coordinates $\alpha = 12^h10^m06^s.3,~ \delta=+18\degree 32'33''.5, ~J2000$ and $l=252.85\degree,~ b=+77.19\degree$ (Harris, 1996; 2010 edition).
Its position overlaps with the tidal stream of the Sagittarius dwarf spheroidal (Sgr dSph) galaxy, a Milky Way satellite that is currently being disrupted, and has been suggested that the cluster is of extragalactic origin (Bellazzini et al., 2003a, b). The cluster has been classified as an Oosterhoff of intermediate type (Oo-Int) (Arellano Ferro et al., 2018a).

Presently there are 19 known variable stars in NGC 4147 listed in the Catalogue of Variable Stars in Globular Clusters (Clement et al., 2001), 15 of which are confirmed RRLs.
In a recent paper (Arellano Ferro et al., 2018a), the RRLs population of the cluster has been studied based on new accurate $VI$
photometry; pulsating periods and fundamental physical parameters were obtained. The latest time-series for these star population was obtained in 2012 and the data are available in electronic form in that paper.

In order to study secular period variations in RRLs, a long time-base is required by any method employed. In particular for the $O-C$ diagram approach, good quality light curves are necessary, in which an accurate estimate of the time of maximum light can be performed. This condition may in fact prove to be a strong limitation when dealing with very old data, when the photometric accuracy was a drawback, particularly for stars in crowded regions such as in globular clusters.

We have searched the literature and compiled data for NGC 4147 from 1954, which means we could have a time-base of about 58 years, which we believe is substantial to figure out the long-term behaviour of the pulsation period of RRLs. The data sources and the
bands used are listed in Table \ref{tab:observations}.

Although the first variable stars in NGC 4147 were discovered early in the XXth century (V4 Davis, 1917; V1-V3 Baade, 1930), the light curves from these works are not available. It was until the work of Newburn (1957), where the variables V5-V14 were discovered, that the first light curves, observed in 1954, became public. Photometric time series from 1955 to 1957 where later published by Mannino (1957) and a complementary discussion on their periods (Mannino, 1958). 

Nearly 45 years elapsed without the variables in NGC 4147 being studied. Arellano Ferro et al. (2004) found a new variable (V18), revised the period and confirmed the non-variable nature of V5, V9 and V15. The observations were conducted for 23 nights in 2003. 

Stetson et al. (2005) presented data from observations made between 1983 and 2003. With the exception of one observation run made by the authors, the rest were donated to the authors or taken by them from public archives. Their work includes light curves and new period estimates. They discovered star V19.

The latest time-series comes from the observations published by Arellano Ferro et al. (2018a), obtained during 8 nights in 2012. Hence, the total time-span of the available data is 58 years.

\MakeTable{lcc}{12.5cm}{Sources of photometric data for the RR Lyrae stars in NGC 4147.\label{tab:observations}}
{
\hline
Authors & Years & Band  \\
\hline
Newburn (1957) & 1954-1955 & $B_{pg}$, $V_{pg}$   \\
Mannino (1957) & 1955-1957 & $pg$   \\
Arellano Ferro et al. (2004) & 2003 & $V$   \\
Stetson et al. (2005) & 1983-2003 & $BVRI$    \\
Arellano Ferro et al. (2018) & 2012 & $V$    \\
\hline
}
\section{The {\it O-C} method}

The observed minus calculated ($O-C$) method has been for many years a very useful approach to determine whether there is a miscalculation in the period of a given star or a real variation in its pulsation period. This method relies on calculating the difference between the observed time of maximum light ($O$) and the calculated one ($C$) given a fixed ephemeris, and then studying the secular behaviour of the residuals ($O-C$).
In fact, one can use any feature of the phased light curve, but in our case, as we are dealing with RRLs, at least with those of the RRab subclass, the maximum of the light curve is well defined, contrary to the minimum that is usually wider.

Taking a given epoch as reference ($E_0$), the calculated time of maximum is given by:

\begin{equation}
C=E_0+P_0N,
\label{eq:calc}
\end{equation}

\noindent
where $N$ is the number of cycles elapsed between $E_0$ and $C$ and $P_0$ is the period at $E_0$. Thus, the number of cycles between $E_0$ and $O$ is,

\begin{equation}
N=\lfloor \frac{O-E_0}{P_0} \rfloor,
\end{equation}

\noindent
where the incomplete brackets indicate the rounding down to the nearest integer.
With this, we can construct the so called $O-C$ diagram by plotting the number of cycles $N$ versus the $O-C$ residuals. The distribution of those points can be modeled by a quadratic distribution:

\begin{equation}
O-C=A_0+A_1N+A_2N^2,
\label{eq:OC}
\end{equation}

\noindent
and, if we substitute eq. \ref{eq:calc} in eq. \ref{eq:OC}, the observed time of maximum is given by

\begin{equation}
O=(A_0+E_0)+(A_1+P_0)N+A_2N^2 .
\label{eq:obs 2}
\end{equation}

The term $A_0+E_0$ is the corrected epoch for the fit, and the term $A_1+P_0$ is the corrected period at that epoch. One can calculate the period at a given $N$ as:

\begin{equation}
P(N)=\frac{dO}{dN}=A_1+P_0+2A_2N.
\label{eq: period at N}
\end{equation}

Since the time elapsed at a given epoch is $t=PN$, the variation of the period with respect of time is

\begin{equation}
\beta\equiv\dot{P}=\frac{dP}{dt}=2A_2\frac{dN}{dt}=\frac{2A_2}{P}.
\label{eq: beta}
\end{equation}

In case we are dealing with a linear distribution of ($O-C$), $A_2=0$ and, as can be seen in eq. \ref{eq: period at N}, we must correct the period $P_0$ to $P_0+A_1$.

\MakeTable{lccc}{12.5cm}{Observed times of maximum light $O$ for the RRL stars in NGC 4147 and the corresponding $O-C$ residual given $P_0$ and $E_0$. \label{tab:t.of max}}{
\hline
Variable & $P_0$(d) & $E_0$(HJD) & \\ 
& & & \\
V1 & 0.500393 & 2455988.49483 & \\
& & & \\
$O$ (HJD) & $O-C$(d) & No. of cycles & Source \\
\hline
2434805.9870		&	-0.3779	&	-42331	&	Nb	\\
2434805.9960		&	-0.3689	&	-42331	&	Nb	\\
2435187.7760		&	-0.3888	&	-41568	&	Nb	\\
2435187.7860		&	-0.3788	&	-41568	&	Nb	\\
2435215.3680		&	-0.3184	&	-41513	&	Mn	\\
2435538.5340		&	-0.4063	&	-40867	&	Mn	\\
2435991.3850		&	-0.4110	&	-39962	&	Mn	\\
2449824.6162		&	-0.0448	&	-12318	&	St	\\
2449825.6034		&	-0.0584	&	-12316	&	St	\\
2449825.6060		&	-0.0558	&	-12316	&	St	\\
2450225.4268		&	-0.0490	&	-11517	&	St	\\
2452761.4621		&	-0.0056	&	-6449	&	St	\\
2452796.4730		&	-0.0222	&	-6379	&	AF04	\\
2455988.5024		&	0	&	0	&	AF18	\\

\hline
\\
\multicolumn{4}{p{7cm}}{The source for the data employed to calculate the times of maximum are coded as follows; Nb: Newburn (1957); Mn: Mannino (1957); St: Stetson et al. (2005); AF04: Arellano Ferro et al. (2004); AF18: Arellano Ferro et al. (2018a). This is an extract from the full table, which is available in electronic format.}
}

\section{The {\it O-C} Diagrams}
\label{OCs}

\begin{figure}
\begin{center}
	\includegraphics[width=12.5cm]{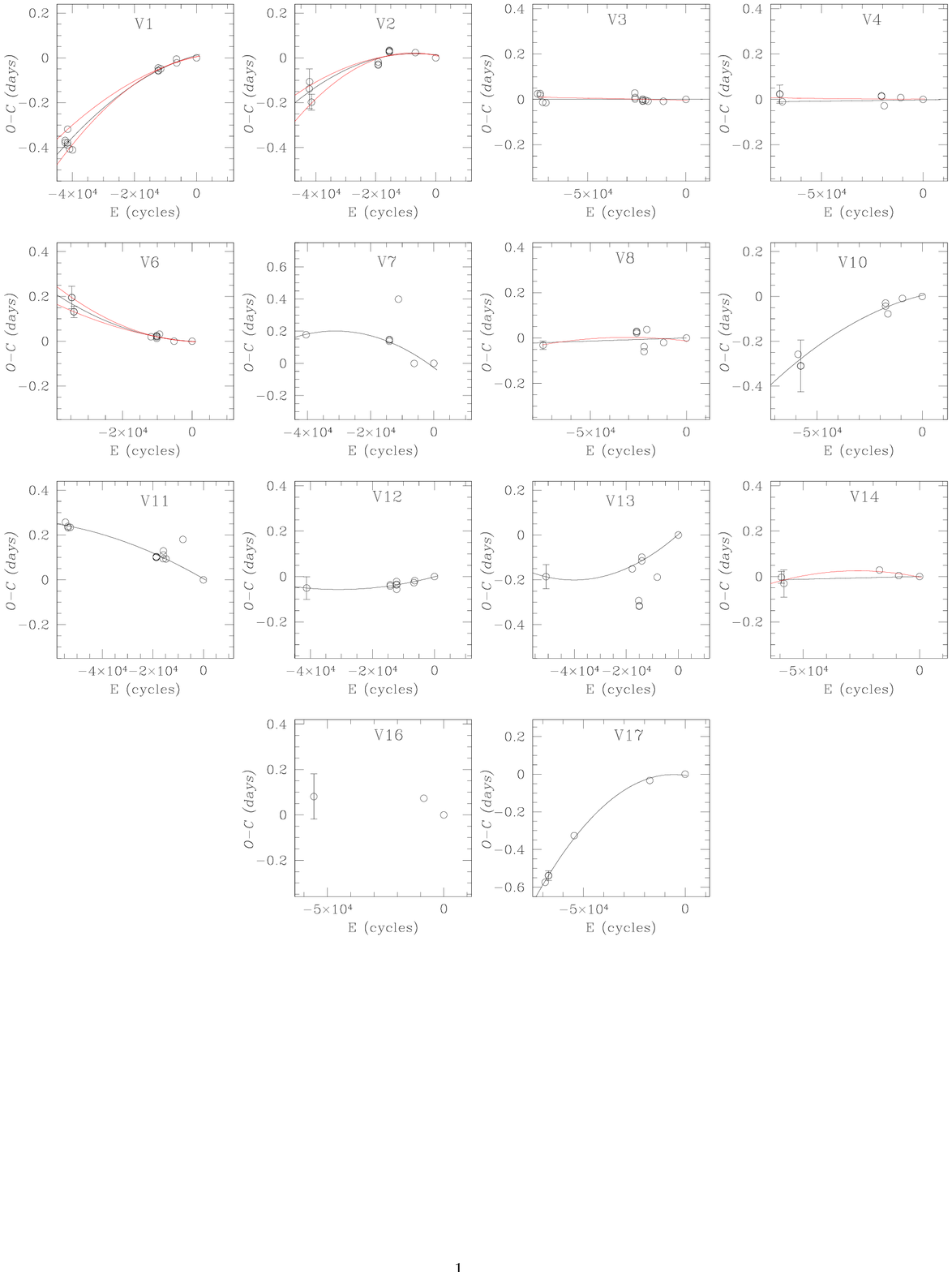}
    \FigCap{$O-C$ diagrams of the RR Lyrae stars in NGC 4147. Points that obviously stand far out of the general trend have not been used in the fits (see V7,V11, V13). Error bars are visible only for some of the older data and when not drawn, their size is smaller than the symbols. When two equally possible solutions are foreseen, two colours are used for the lines. See $\S$ \ref{OCs} for a detailed discussion.}
    \label{fig:o-c}
\end{center}
\end{figure}

Table \ref{tab:t.of max} lists the times of maximum observed for each RRL, with the correspondent $O-C$ residuals, number of cycles elapsed and the source from which we obtained the data; only an extract for V1 of the full table is shown, but the complete table is available in electronic form. The ephemerides ($P_0$, $E_0$) used to calculate the $O-C$ residuals were taken from Arellano Ferro et al. (2018a, Tab. 3), and are also listed in Table \ref{tab:gral} of this paper. A plot of column 2 vs. column 3 of Table \ref{tab:t.of max} produces the $O-C$ diagram of each star.
In Fig. \ref{fig:o-c} the individual $O-C$ diagrams are displayed. Uncertainties in the determination of the observed times of maximum are particularly large in older data due to the scatter of the light curves. Error bars are plotted. For more recent
data, the uncertainties are small and the corresponding error bars are
comparable to the size of the symbols and are thus not plotted. When two competing solutions are possible, red and black lines were used to distinguish them.

\MakeTable{llccc}{12.5cm}{General data for the RRL variables in NGC 4147.\label{tab:gral}}{
\hline
Variable & Variable & $P_0$ & $E_0$ (HJD) & $\beta$ ($O-C$) \\
Star ID & Type & (days) & (+2 450 000) & (d Myr$^{-1}$) \\ 
\hline

V1 & RRab Bl & 0.500393 & 5988.4948 & -0.196 $\pm$ 0.025 \\
V2 & RRab Bl & 0.493297 & 5988.3470 & -0.202 $\pm$ 0.019 \\
V3 & RRc & 0.280543 & 5986.5190 & -0.001 $\pm$ 0.003 \\
V4 & RRc & 0.300066 & 5963.2723 & +0.004 $\pm$ 0.007 \\
V6 & RRab Bl & 0.609737 & 5963.3748 & +0.139 $\pm$ 0.020 \\
V7 & RRab Bl& 0.514321 & 5986.4396 & -0.333 $\pm$ 0.034 \\
V8 & RRc & 0.278599 & 5963.4587 & -0.044 $\pm$ 0.010 \\
V10 & RRc & 0.352339 & 5963.2378 & -0.092 $\pm$ 0.014 \\
V11 & RRc & 0.387423 & 5963.3836 & -0.076 $\pm$ 0.007 \\
V12 & RRab & 0.504700 & 5986.5132 & +0.086 $\pm$ 0.008 \\
V13 & RRc & 0.408540 & 5963.3660 & +0.225 $\pm$ 0.010 \\
V14 & RRc & 0.356375 & 5963.4590 & -0.082 $\pm$ 0.010 \\
V16 & RRc & 0.372134 & 5963.4590 & -- \\
V17 & RRc & 0.374843 & 5988.3090 & -0.276 $\pm$ 0.006 \\
\hline
}

Except for a few divergent points in some stars, for example in V7, V11 and V13, the rest of the $O-C$ residuals follow a coherent trend and suggest in most cases a mild period variation.

In V1, V2, and V6 a clear period variation is displayed in the $O-C$ diagrams of these stars. Two possible solutions are indicated since a larger scatter is seen in the older data. In all cases the difference in the value of $\beta$ for the two extreme solutions is about 0.1 d Myr$^{-1}$. The finally adopted value of $\beta$ comes from the intermediate solution shown as a black parabola.

In V3, V4, V8, and V14, a rather linear and horizontal $O-C$ distribution is observed, meaning that the assumed ephemeris period $P_0$ is constant and nearly correct. Even in this cases we have fitted a parabola to estimate an upper limit on $\beta$.

For V16 only three reliable times of maximum light could be found, therefore its solution is uncertain. 
The values of $\beta$ and their uncertainties are listed in Table \ref{tab:gral}.

V18 is not considered as it is a semi-regular red variable confirmed as such by  Arellano Ferro et al. (2018a).
The RRab star V19 was discovered by Stetson et al. (2005) only in 2005 and data do not span sufficient time, hence it was not included in this work.

\section{Observed mean secular period variations in the HB and the theoretical predictions}

\begin{figure}
    \begin{center}
	\includegraphics[width=8.8cm]{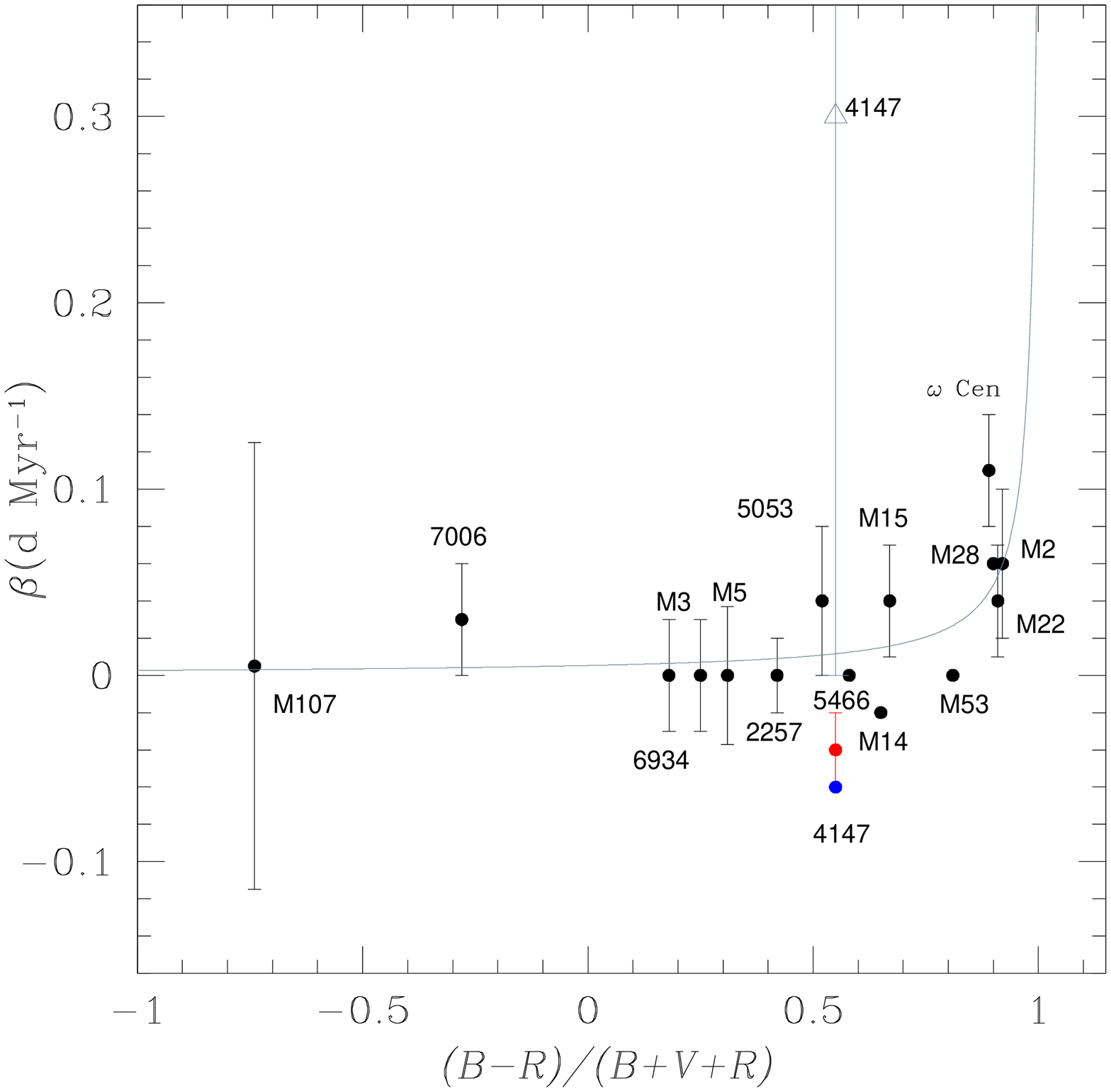}
    \FigCap{Median period change rate as a function of the HB structure parameter $\mathcal{L}\equiv (B-R)/(B+V+R)$. The numbers correspond to the NGC nomenclature. The position of NGC~4147 shown with a red dot corresponds to the median calculated without considering the Blazhko variables. If these are included the position at the blue symbol is obtained. The outlying position of NGC 4147 marked with an open triangle (Catelan, 2009) is not supported by the present results. See $\S 5$ for a discussion.}
    \label{fig:beta lee-zinn}
    \end{center}
\end{figure}

It has been argued in several works that the mean RRLs period variations, in globular clusters should be close to zero, although some increase in $\beta$ is expected for clusters with very blue HB's, i.e. with large values of the Lee-Zinn structure parameter $\mathcal{L}$ ($\equiv (B-R)/(B+V+R)$, where $B$, $V$, and $R$ are the number of stars to the blue, inside, and to the red of the
instability strip (IS), respectively), as predicted by theoretical models (Lee, 1991; Catelan, 2009). We shall refer the reader to Fig. 15 in Catelan (2009) or to its updated version in Fig. \ref{fig:beta lee-zinn} of the present paper. However, as stressed by Smith (1995) and Catelan (2009), the observational trend rests on $\omega$ Cen. The rest of the clusters display a dispersion of $\pm$ 0.05 d Myr$^{-1}$ around zero. 
We have to recognize however, that the empirical determinations of $\beta$ considered in the figure are of very inhomogeneous quality, the main reason being that long time-span of observations of fair quality is required and these conditions are not always met. There is a rather small number of clusters that satisfy these requirements; they are listed in $\S$ 1.

In Fig. \ref{fig:beta lee-zinn} the solid line is the theoretical prediction, represented mathematically by equation 6.5
of Catelan \& Smith (2015).
Black dots represent the data from Catelan (2009, Tab. 4), collected from empirical works in the literature, with the exception of M5 ($<\beta>=0.0 \pm 0.04$) and M107 ($<\beta>=+0.005 \pm 0.12$), for which the results were taken from Arellano Ferro et al. (2016) and Arellano Ferro et al. (2018b) respectively. The triangle showing the locus of NGC 4147 was calculated by Catelan (2009) based on data from Stetson et al. (2005). Its position stands out and triggers doubts on the median $\beta$ for this cluster. To some extent the present work was encouraged by this unexpected result for NGC 4147.

The median of all the values of $\beta$ reported in Table \ref{tab:gral} is $\beta=-0.06 \pm 0.02$ d Myr$^{-1}$. While this value places the cluster much closer to the theoretical locus, it is in fact a bit too negative, although it could be argued that, given the uncertainties, it lies within the scatter displayed by the rest of the sample. The negative median is produced mostly by the presence in the sample of a few stars with significant negative values of $\beta$; V1, V2, V7 and V17. A close inspection of the light curves of these stars in the work of  Arellano Ferro et al. (2018a), reveals clear signs of Blazhko modulations in V1, V2 and V7, as well as in V6, which has a significant positive value of $\beta$. Since the presence of amplitude and phase modulations do introduce uncertainty in the $O-C$ diagram, we also calculated the median by eliminating the four Blazhko variables listed in Table \ref{tab:gral}, to find the value $\beta=-0.04 \pm 0.02$ d Myr$^{-1}$. In either case it is clear that the median period changes in NGC~4147 are close to zero, as expected from theoretical arguments.
 
Stars with significant negative values of $\beta$, hence evolving to the blue across the IS, are always a matter of discussion. According to the post-ZAHB models of Dorman (1992), the fastest blueward evolution, in the loop very near to the ZAHB, reaches the rate -0.026 d Myr$^{-1}$. It has been argued that larger negative period change rates can be found if pre-ZAHB stars, on their contraction to the ZAHB, cross the IS disguised as RRL star (Silva Aguirre et al., 2008). Nevertheless, according to the simulations made by these authors for M3, it is very unlikely that a pre-ZAHB falls in the IS (see also the discussion of Arellano Ferro et al. (2018b) in their section 5 for the case of M107). In spite of this, there are numerous stars in several globular clusters with well determined negative values, with $\beta$ $< -0.3$ d Myr$^{-1}$; five in M3 (Corwin \& Carney, 2001), five in M5 (Arellano Ferro et al., 2016), four in NGC6934 (Stagg \& Wehlau, 1980) and three in M107 (Arellano Ferro et al., 2018b).
 
Finally, we want to pay attention to the fact that stochastically irregular period variations are not found in any of the RRL stars in NGC~4147 as they were also not found in the RRL population of M5 (Arellano Ferro et al., 2016) and of M107 (Arellano Ferro et al., 2018b).
It was demonstrated by Arellano Ferro et al. (2016) that a miscounting of cycles can produce odd-looking $O-C$ diagrams which can be erroneously interpreted as due to stochastic period variations. This clarification is correct since it is clear from the above studies that, with very few exceptions, regular $O-C$ diagrams can be obtained, even in extreme Blazhko variables, if a proper cycle counting is performed. This statement does not contradict the theoretical arguments for stochastic period variations induced by mixing events in the core that may alter the hydrostatic structure and the pulsation period (e.g. Bal\'azs-Detre \& Detre, 1965; Sweigart \& Renzini, 1979), but rather stresses the need of long-term continuous observations to detect period irregularities as those observed in RR Lyrae itself (Le Borgne et al., 2008). In other words, it is possible that some of the regular secular period variations, as those reported in the present work, co-exist in combination of secondary variations, of stochastic nature, only detectable in dense favourable time-series.

\section{Summary}

Photometric data of RRL stars in NGC 4147 were gathered from the literature, spanning 58 years. They were employed to estimate secular period variations $\beta$, in 13 stars of both RRab and RRc type, via the traditional $O-C$ approach. Given the HB structure parameter $\mathcal{L} = 0.54$, the expected median value of $\beta$ according to theoretical predictions should be close to zero. In fact, the median found in this work (excluding Blazhko variables) is $\beta=-0.04 \pm 0.02$ d Myr$^{-1}$, which is not inconsistent with theory if the scatter in the empirically determined median values of $\beta$ is considered. It must be admitted however, that empirical results shown in Fig. \ref{fig:beta lee-zinn} come from numerous sources and are subject to large inhomogeneities, both from the methodology and the quality of available data. Error bars are necessarily large due to the presence of stars in the samples  with authentically large period change rates, both positive and negative. Scatter and error bars are simply too large to meet a satisfactory comparison with theory. Noticeably, clusters with the bluest HB, M2, M22 and M28 do not significantly differ from the rest of the clusters, whereas
 $\omega~Cen$, with a slightly redder HB stands out for having a large median $\beta$, which perhaps only contributes to show the uniqueness of this cluster. It seems appropriate to join on the conclusion reached by Smith (1995) more than 20 years ago: "The data at hand are broadly consistent with [theoretical] predictions, but are not adequate to provide confirmation".

\Acknow{
A.L. is indebted to CONACyT M\'exico for the graduate scholarship [fellowship number 672829].
The present research was partially financed by DGAPA-UNAM, M\'exico [grant number IN106615-17] for which we are grateful. We are grateful to Prof. Sunetra Giridhar for her comments on the manuscript. We have made extensive use of the SIMBAD and ADS services.}

\end{document}